\documentclass[reprint, superscriptaddress, fleqn, amsmath,amssymb, aps, floatfix, nofootinbib]{revtex4-2}
\usepackage{amsmath}
\usepackage{graphicx}% Include figure files
\usepackage{dcolumn}% Align table columns on decimal point
\usepackage{bm}% bold math
\usepackage[LGRgreek]{mathastext}
\usepackage{makecell}
\usepackage{multirow}
\usepackage{hyperref}
\hypersetup{colorlinks=true, citecolor=blue, urlcolor=blue, linkcolor=blue}
\newcommand{\isotope}[2]{\textsuperscript{#2}#1}

\newcommand{\ta}[1]{\textsuperscript{#1}Ta}
\newcommand{\hf}[1]{\textsuperscript{#1}Hf}

\begin{document}

\title{\ta{179}(n,$\gamma$) cross-section measurement and the astrophysical origin of \ta{180} isotope}

\author{R. Garg}
\email[Corresponding email: ]{ruchi.garg.phys@gmail.com}
\affiliation{School of Physics and Astronomy, University of Edinburgh, Edinburgh, EH9 3FD, UK}
\affiliation{Facility for Rare Isotope Beams, Michigan State University, East Lansing, Michigan, 48824, USA}
\author{S. Dellmann}
\affiliation{Goethe Universit$\ddot{a}$t Frankfurt, Frankfurt, 60438, Germany}
\author{C. Lederer-Woods}
\affiliation{School of Physics and Astronomy, University of Edinburgh, Edinburgh, EH9 3FD, UK}
\author{C. G. Bruno}
\affiliation{School of Physics and Astronomy, University of Edinburgh, Edinburgh, EH9 3FD, UK}
\author{K. Eberhardt}
\affiliation{Johannes Gutenberg Universit$\ddot{a}$t Mainz, Mainz, 55128, Germany}
\author{C. Geppert}
\affiliation{Johannes Gutenberg Universit$\ddot{a}$t Mainz, Mainz, 55128, Germany}
\author{T. Heftrich}
\affiliation{Goethe Universit$\ddot{a}$t Frankfurt, Frankfurt, 60438, Germany}
\author{I. Kajan}
\affiliation{Paul Scherrer Institute, Villigen, 5232, Switzerland}
\author{F. K\"appeler}
\altaffiliation{Deceased}
\affiliation{Karlsruhe Institute of Technology, Campus North, Karlsruhe, 76021, Germany}
\author{B. Phoenix}
\affiliation{School of Physics and Astronomy, University of Birmingham, Birmingham, B15 2TT, UK}
\author{R. Reifarth}
\affiliation{Goethe Universit$\ddot{a}$t Frankfurt, Frankfurt, 60438, Germany}
\author{D. Schumann}
\affiliation{Paul Scherrer Institute, Villigen, 5232, Switzerland}
\author{M. Weigand}
\affiliation{Goethe Universit$\ddot{a}$t Frankfurt, Frankfurt, 60438, Germany}
\author{C. Wheldon}
\affiliation{School of Physics and Astronomy, University of Birmingham, Birmingham, B15 2TT, UK}

\begin{abstract}
\ta{180m} is nature's rarest (quasi) stable isotope and its astrophysical origin is an open question. A possible production site of this isotope is the slow neutron capture process in Asymptotic Giant Branch stars, where it can be produced via neutron capture reactions on unstable \ta{179}. We report a new measurement of the \ta{179}($n,\gamma$)\ta{180} cross section at thermal neutron energies via the activation technique. Our results for the thermal and resonance-integral cross-sections are 952 $\pm$ 57 b and 2013 $\pm$ 148 b, respectively. The thermal cross section is in good agreement with the only previous measurement (Phys. Rev C {\bf 60} 025802, 1999), while the resonance integral is different by a factor of $\approx$1.7. While neutron energies in this work are smaller than the energies in a stellar environment, our results may lead to improvements in theoretical predictions of the stellar cross section. 
 
\end{abstract}

\maketitle

%======================= INTRODUCTION =======================
\section{Introduction}
\begin{figure}[hb!]
    \centering
    \includegraphics[width=0.80\linewidth]{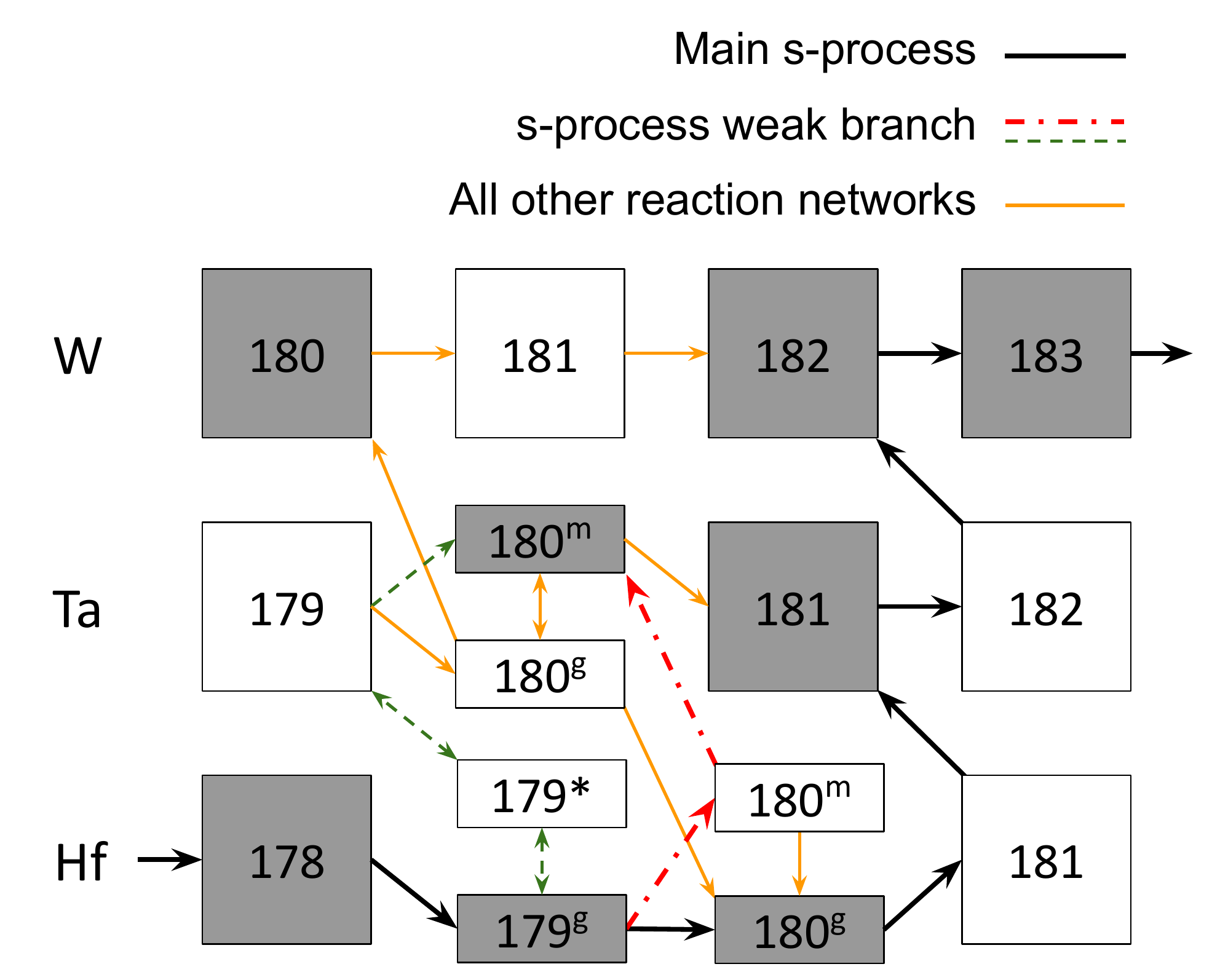}
    \caption{s-process reaction network. Grey and white boxes show the stable and unstable isotopes respectively. Thick black arrows show the main s-process path. The red and green arrows show the weak branching paths suggested by Refs. \cite{Beer81} and \cite{Yokoy83} respectively. The orange arrows show all the other reactions in the network.}
    \label{fig:rxn_net}
\end{figure}

Tantalum-180 is one of the most interesting isotopes in nature. In its ground state, this isotope is unstable with a half-life of 8.15 hours, however, it has a high-spin (9$^{-}$) meta-stable state at 77.2~keV that has a half-life of $>$ 7 $\times$ 10$^{15}$~years. This isomer, \ta{180m}, is nature's rarest (quasi) stable isotope and its stellar origin remains an open question. At least 3 nucleosynthesis processes are thought to contribute to the \ta{180} abundance. References \cite{Woosley90, Heger05, Hayakawa10, Sieverding18} suggest that \ta{180} is produced by \hf{180} ($\nu_e$,e)\ta{180} and \isotope{Ta}{181}($\nu,\nu$'n)\ta{180} reactions in the $\nu$-process in stellar explosions. Another proposed site is the p-process in O/Ne-rich layers in Type-II supernovae (SNII), where the ($\gamma$,n) reactions on \ta{181} lead to \ta{180} production~\cite{Prantzos90, Rayet95, Utsunomiya03, Malatji19}. Finally, in low mass asymptotic giant branch (AGB) stars, two reaction sequences have been suggested as source for \ta{180}: (i) neutron capture on \hf{179}  resulting in an isomeric  state of \hf{180} (J$^\pi=8^-$, 1141~keV), which has a small $\beta$ decay branch to \ta{180m}\, and (ii) $\beta$ decay of thermally excited states in \hf{179} to \ta{179}, and subsequent neutron capture to \ta{180m} \cite{Yokoy83}. Figure~\ref{fig:rxn_net} shows the two reaction paths with red and green arrows respectively. K\"appeler \textit{et al.}~\cite{Kappeler04} estimate that (ii) can explain 80--86\% of the solar \ta{180} abundance, while  path (i) seems to only contribute to a small extent \cite{Kellog92}. However, a recent study~\cite{Malatji19} modelled s-process nucleosynthesis in AGB stars using the neutron capture cross-sections derived from statistical models (using experimentally obtained nuclear structure parameters~\cite{Brits19}), and found only a negligible contribution to the observed \ta{180} abundance. They also studied the impact of the newly constrained value of the \ta{179}(n,$\gamma$) cross-sections on the time reversed reaction, \ta{180}($\gamma$,n)\ta{179}, which is the main mode of destruction of \ta{180m} in the SNII p-process. They found that the new reaction rate reduces the \ta{180m} overabundance in the p-process models. The variety of different predictions emphasises the need for accurate experimental data on nuclear reactions and stellar half lives for the isotopes involved.  

The destruction reaction \ta{180m}(n,$\gamma$) has been measured by Wisshak \textit{et al.}~\cite{Wisshak01}. A direct measurement of \ta{179}(n,$\gamma$) cross-section at neutron energies relevant to s-process temperatures (keV neutron energies) has not been possible yet, due to the lack of availability of a radioactive \ta{179} (T$_{1/2}$ = 1.82 y) target with sufficient mass. However, the larger neutron fluxes available at research reactors allow an activation measurement of the \ta{179}($n,\gamma$)\ta{180} reaction at thermal neutron energies (25 meV). There is only one previous measurement of the thermal \ta{179}($n,\gamma$)\ta{180} cross section and resonance integral~\cite{Schumann99}. This article reports the results of a new measurement of this important reaction cross section.
    
%======================= METHOD =======================

\section{Method}
    %=======================
    \subsection{Radioactive target production}
    %=======================
    The \ta{179} sample was produced via \hf{180}(p,2n) reactions on a metallic Hf foil, using the same approach as Ref. \cite{Schumann99}. The hafnium foil ($>99.63$\% purity) of 1.331~g was irradiated for $\approx$7~hours with a 27-MeV beam of protons and a current of 32~$\mu$A at the MC40 cyclotron at the University of Birmingham. In addition to the isotope of interest, \ta{179}, the proton irradiation also produced a few other radioactive isotopes in the sample, the most dominant ones being $^{172, 175, 181}$Hf and $^{172}$Lu. After a cool-down period of $\approx$7 months, the sample was treated at PSI for the radio-chemical separation of \ta{179} as follows: The obtained Hf foil was partially dissolved in concentrated hydrofluoric acid (HF) with a few drops of HNO$_{3}$. An aliquot of the sample was measured by a High-Purity Germanium (HPGe) detector to identify the main radioactive isotopes in the sample. The most dominant $\gamma$-lines were attributed to \isotope{Hf}{175} and \isotope{Hf}{172}, together with its radioactive daughter \isotope{Lu}{172}. The hafnium isotopes were then conveniently used as tracers to follow the efficiency of the chemical separation of \ta{179}. After the dissolution, the sample was diluted by Milli-Q water, changing the HF concentration to 1~Mol·L$^{-1}$. The solution was then applied on a chromatographic column filled with TBP resin (Triskem). The majority of the hafnium in the solution passed through the column whereas \ta{179} was absorbed on the resin. Afterwards, \ta{179} was eluted using 0.1~M HCl solution. The decontamination factor, Df, from hafnium was determined using,
    \begin{equation}
        \label{eqn:sep}
        Df = \frac{C_{Hf}^0}{C_{Hf}^s} \thickapprox \frac{A_{^{175}Hf}^0}{A_{^{175}Hf}^s},
    \end{equation}
    where $C^0$ and $C^{s}$ represent relative hafnium concentrations before and after the separation respectively, while $A^0$ and $A^s$ represent the radioactivity of \isotope{Hf}{175} demonstrated by its 343.4~keV $\gamma$-ray emission from the sample before and after the separation, respectively. The decontamination factor Df was measured to be of the order of 300.

    The solution containing \ta{179} was afterwards evaporated to near dryness and re-dissolved in a mixture of 6 Mol·L$^{-1}$ HCl/20 mMol·L$^{-1}$ HF. The solution was then applied to a chromatographic column containing 2~g of TEVA resin as described by Snow \textit{et. al.}~\cite{Snow_17}. The \ta{179} was absorbed in the column whereas hafnium was eluted as the column was washed with a fresh acid mixture. After no signal from \isotope{Hf}{175} could be detected with a HPGe detector, the \ta{179} was eluted from the column using 6 Mol·L$^{-1}$ HNO$_3$/20 mMol·L$^{-1}$ HF mixture. After the $\gamma$-spectroscopic analysis only a weak signal of \isotope{Hf}{175} was found in the eluate containing \ta{179}, yielding a minimum separation factor on the order of 1$\times10^4$. The separation procedure on TEVA resin was repeated once more without use of additional radiotracers in order to remove the last traces of hafnium from the sample. The extracted sample was sealed in the tip of an Eppendorf vial for transportation to Mainz for the activation at the TRIGA reactor.
    
    The activity of the extracted sample, as measured at Mainz on day 521 from the proton irradiation of the hafnium foil, was found to be 1.905(55)~MBq. This activity corresponds to a \ta{179} mass of 47~ng or 1.58$\times10^{14}$ atoms. The activity was determined by measuring \ta{179} X-rays \cite{Baglin09} at energies 54.61 and 55.79~keV of intensities 12.6(3) and 21.8(5)\% respectively, on two identical low energy HPGe detectors. The uncertainty on the measured activity was calculated by combining the uncertainties of the X-ray intensities, detection efficiencies, and the statistical error on the peak counts.

    %=======================
    \subsection{Reactor activation and Cd difference method}
    %=======================
    The \ta{179}($n,\gamma$)\ta{180} cross section was measured via the activation technique, which consists of exposing the sample to a flux of neutrons and subsequent counting of the reaction product \ta{180} through its radioactive decay. This allows the neutron capture cross section to be determined using 
    \begin{equation}
    \label{eqn:act_simple}
        \sigma \Phi = \frac{N_{\ta{180}}}{N_{\ta{179}}},
    \end{equation}
    where $N_{\ta{179}}$ and $N_{\ta{180}}$ is the number of \ta{179} and \ta{180}  nuclei, respectively. The neutron fluence (time integrated flux), $\Phi$, was determined by irradiation of isotopes with well known cross sections (table~\ref{tab:mon_iso}) and $N_{\ta{180}}$ was determined by decay counting of its 93.3~keV gamma-line.  It should be emphasised here that since the activation technique can only be used for reactions with a radioactive product, the present work only determines the cross section to the unstable ground state of \ta{180}. 
    
    Irradiation of the \ta{179} sample was performed at the TRIGA reactor in Mainz, in a rotating irradiation carousel, which results in exposure to a uniform neutron flux. The reactor's neutron energy spectrum is comprised of 3 main components: moderated thermal neutrons with a Maxwell-Boltzmann distribution corresponding to $kT=25.3$ meV, epithermal neutrons (0.2 eV -- 0.5 MeV) with the flux exhibiting a 1/E energy dependence, and fast  fission neutrons (E$_n >$ 0.5 MeV). For the present experiment, the contribution from fast neutrons to the cross section is negligible due to the low flux and low reaction cross sections.
    
    In the presence of these different components of the neutron flux, the activation equation becomes,
    \begin{equation}
        \label{eqn:act_gen}
        \sigma_{th}^{\isotope{X}{A}}~\Phi_{th} + I_{res}^{\isotope{X}{A}}~\Phi_{epi} = \frac{N_{\isotope{X}{A+1}}}{N_{\isotope{X}{A}}},
    \end{equation}
    where, $\Phi_{th}$ and $\Phi_{epi}$ are the thermal and epithermal components of the neutron fluence, $\sigma_{th}^{^{A}X}$ is the cross-section for a thermal neutron-capture on the \isotope{X}{A} nucleus, $I_{res}^{\isotope{X}{A}}$ is the resonance integral of the cross-section in the epithermal energy region, and $N_{\isotope{X}{A}}$ and $N_{\isotope{X}{A+1}}$ are the number of target nuclei and the activated nuclei respectively.
    
    To measure the contribution to the activation of the sample from the thermal and epithermal components of the fluence, the cadmium difference method was employed. This involved two irradiations of the $^{179}$Ta sample, one with, and one without a surrounding cadmium shield of 1~mm thickness, which shields the sample from thermal neutrons due to the high thermal absorption cross section of \isotope{Cd}{113}. Each irradiation was performed for 3 hours, 20 days apart to ensure that no  $^{180}$Ta was left in the sample before the second irradiation.
    
    The neutron fluence ($\Phi_{th}$ and $\Phi_{epi}$) that the \ta{179} sample was exposed to during the activations, was measured using a combination of isotopes with well known neutron-capture cross-sections at thermal and epithermal energies. Two samples of each monitor isotope were placed on opposite sides of the \ta{179} sample during activations, allowing for the determination of the fluence at the \ta{179} location by averaging. The fluence monitor samples that were used in the two activations are listed in table~\ref{tab:mon_mass}. Table~\ref{tab:mon_iso} lists the monitor isotopes along with their abundances and the ($n,\gamma$) cross-sections at thermal and epithermal neutron energies. This combination of isotopes was chosen since the cross-sections are already known with high accuracy and their excitation functions show different behaviours.

    {
    \renewcommand{\arraystretch}{1.1}
    \setlength{\extrarowheight}{1.1pt}
    \setlength{\tabcolsep}{12pt}
    \begin{table}[h]
    \centering
    \begin{tabular}{|c|c|c|}
        \hline
        Activation & Sample  &  Mass (mg)\\
        \hline \hline
        \multirow{4}{*}{1st: No shield} &
           Au-1  &  $(5.91\pm0.01) \times 10^{-3}$\\ \cline{2-3}
           & Au-2  &  $(6.33\pm0.01) \times 10^{-3}$\\ \cline{2-3}
           & Zr-1  &  23.4 $\pm$ 0.1\\ \cline{2-3}
           & Zr-2  &  23.5 $\pm$ 0.1\\
        \hline
        \multirow{6}{*}{2nd: Cd-shield} &
           Au-3  &  $(5.91\pm0.01) \times 10^{-3}$\\ \cline{2-3}
           & Au-4  &  $(6.33\pm0.01) \times 10^{-3}$\\ \cline{2-3}
           & Zr-3  &  23.4 $\pm$ 0.1\\ \cline{2-3}
           & Zr-4  &  23.5 $\pm$ 0.1\\ \cline{2-3}
           & Sc-3  &  1.45 $\pm$ 0.01\\ \cline{2-3}
           & Sc-4  &  1.54 $\pm$ 0.01\\
        \hline
    \end{tabular}
    \caption{Masses of the fluence monitor samples.}
    \label{tab:mon_mass}
    \end{table}
    }
    
    {
    \renewcommand{\arraystretch}{1.1}
    \setlength{\extrarowheight}{1.1pt}
    \setlength{\tabcolsep}{6pt}
    \begin{table}[h]
    \centering
    \begin{tabular}{|c|c|c|c|}
        \hline 
        Isotope & Abundance & $\sigma_{thermal}$ & $I_{res}$ \\
        & (\%) & (b) & (b)\\
        \hline \hline
        $^{45}$Sc & 100 & 27.16 $\pm$ 0.20 & 12.0 $\pm$ 0.05 \\
        \hline
        $^{94}$Zr & 17.4 $\pm$ 0.3 & 0.0498 $\pm$ 0.0017 & 0.265 $\pm$ 0.01 \\
        \hline
        $^{96}$Zr & 2.8 $\pm$ 0.1 & 0.0229 $\pm$ 0.0010 & 5.15 $\pm$ 0.11  \\
        \hline
        $^{197}$Au & 100 & 98.67 $\pm$ 0.09 & 1550.0 $\pm$ 28.0 \\
        \hline
    \end{tabular}
    \caption{List of isotopes used to measure neutron fluence. The cross-section values are taken from the Atlas of Neutron Resonances~\cite{Mughabghab06}.}
    \label{tab:mon_iso}
    \end{table}
    }
    
    %=======================
    \subsection{Gamma-activity measurement}
    %=======================
    \label{sec:setup}

\begin{figure}[h]
        \centering
        \includegraphics[width=0.55\linewidth]{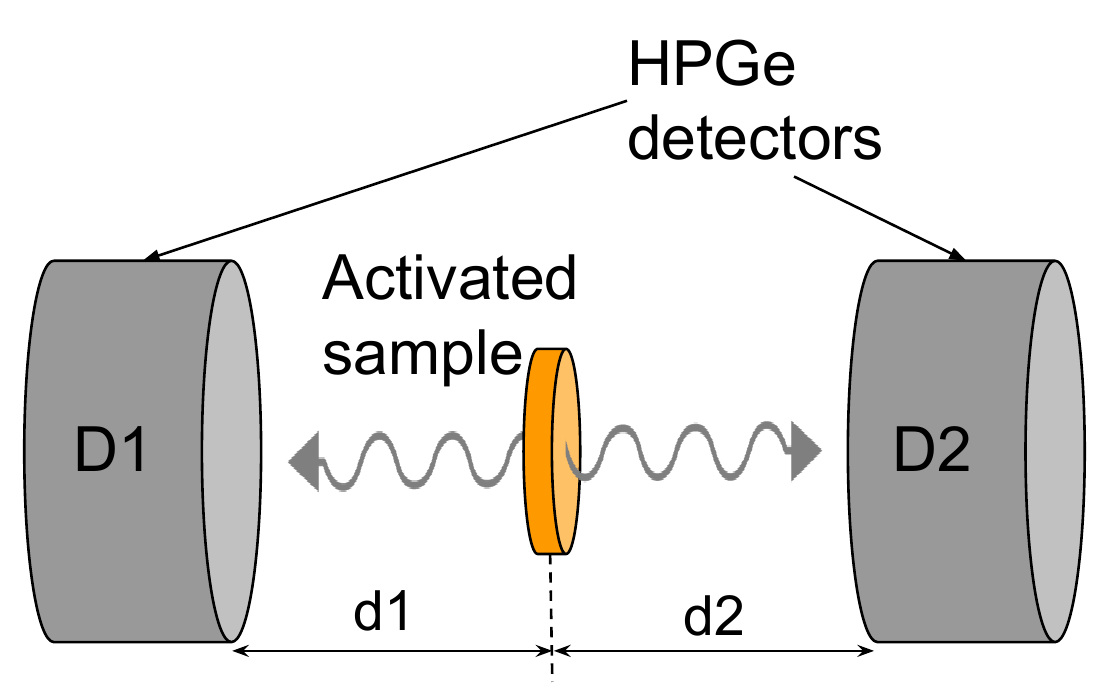}
        \caption{Schematic diagram of the detector setup comprising of two identical HPGe detectors D1 and D2 placed at distances d1 and d2 from the source holder, respectively.}
        \label{fig:detectors}
    \end{figure}

    The gamma-activity of the irradiated samples (\ta{179} and the fluence monitors) was measured using two identical HPGe detectors arranged as shown in figure \ref{fig:detectors}. The sample was secured in a holder and the two detectors were placed at opposite sides of the sample. The whole setup was placed inside lead shielding to reduce background from natural radioactivity. The detectors were placed at 5~cm from the source holder to measure the activity from irradiated \ta{179} sample.
    
\begin{figure}[h!]
        \centering
        \includegraphics[width=1.0\linewidth]{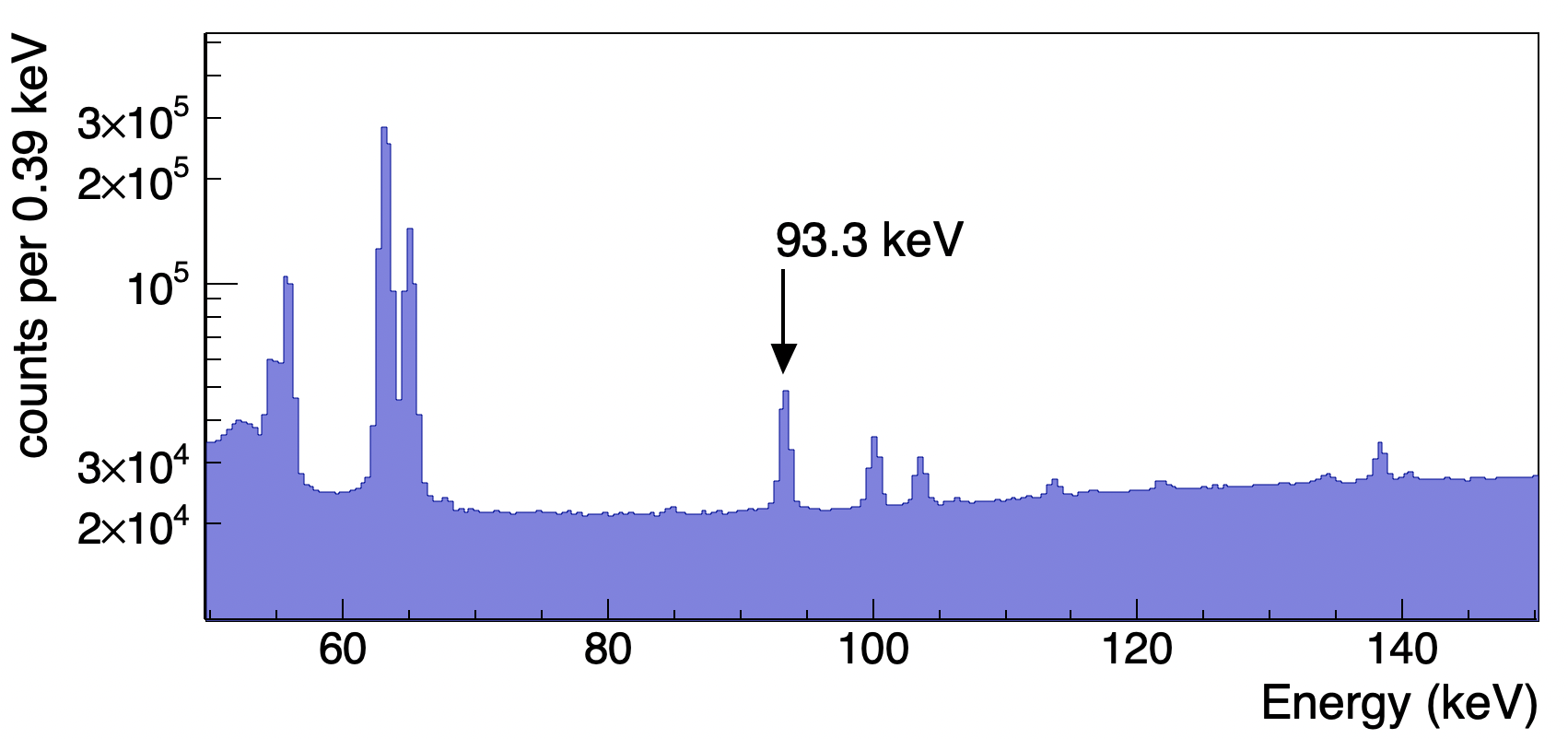}
        \caption{Gamma-ray spectrum of activated \ta{179} as recorded in detector D1 (see figure~\ref{fig:detectors}) over 12 h starting 5.5 h after the irradiation.}
        \label{fig:ta_spec}
    \end{figure}

    Tantalum-180 activity was measured using $\gamma$ rays following the $\beta$-decay of \ta{180} to the first 2$^+$ state of \hf{180} at 93.3~keV (with 4.51\% intensity). At this low energy, the background from 55--65 keV X-rays from \ta{179} was significant. Therefore, an indium absorber disc of 1.5~mm was placed in front of each detector for \ta{180} activity measurements. Figure~\ref{fig:ta_spec} shows a $\gamma$-ray spectrum of the irradiated \ta{179} sample with the 93.3~keV line clearly marked.

    The fluence monitor samples had a higher activity compared to the \ta{179} sample due to their higher mass. Therefore, their activity measurements required the detectors to be placed farther at $\approx$11~cm from the sample in order to reduce the dead time.
    
%======================= Data Analysis =======================

\section{Data analysis and results}

    %=======================
    \subsection{Activation ratio}
    %=======================
    The activation ratio, $R$, is the number of activated nuclei divided by the number of target nuclei. This ratio for an irradiated sample was obtained using the following equation.
    \begin{equation}
        \label{eqn:act_calc}
        R = \frac{C_\gamma}{\epsilon(E_{\gamma}, d) I_\gamma f_a f_w f_m N_{target}} ,
    \end{equation}
    where,
    {\setlength{\mathindent}{0cm}
    \small
    \begin{flalign*}
        C_\gamma &= \text{counts in the characteristic $\gamma$ peak,}\\ 
        \epsilon(\text{\small{E$_{\gamma}$,d}}) &= \text{absolute detection efficiency at ${\gamma}$ energy, E$_{\gamma}$,}\\
        &~~~~\text{and detector distance d,}\\
        I_\gamma &= \text{decay intensity of the characteristic $\gamma$ line,}\\
        f_a &= \text{correction for decay during activation time, }t_a\\
            &= \frac{1-e^{-\lambda t_a}}{\lambda t_a},\\[1ex]
        f_w &= \text{correction for decay during waiting period, }t_w\\
            &= e^{-\lambda t_w},\\
        f_m &= \text{correction for decay during measurement time, }t_m\\
            &= 1 - e^{-\lambda t_m},\\
        \lambda &= \text{decay constant}.
    \end{flalign*}
    \normalsize}
        
    %=======================
    \subsection{Efficiency of the detectors}
    %=======================
    
    To calculate the activation ratio of a sample using equation~\ref{eqn:act_calc}, the absolute detection efficiencies of the detectors were required at the sample's characteristic $\gamma$-ray energies and for the corresponding detection setup arrangement. GEANT3 simulations~\cite{Brun_97} of the setup accurately modelled the detector response as a function of $\gamma$-ray energy, the attenuation from an absorber of given thickness, and the geometrical acceptance of the setup for a given distance between detector and the source. Therefore, to obtain the absolute detection efficiencies, $\epsilon(E_\gamma, d)$, the only unknowns in the simulations were the distances of the detectors to the source-holder (d1 and d2 in fig.~\ref{fig:detectors}).

    To precisely determine the detector-to-sample-holder distances, the simulated efficiencies, $\epsilon(E_{\gamma},d)$, at different $E_{\gamma}$ and d values (varying in the intervals of 1 mm) were compared with the efficiency values of the setup measured using the radioactive sources of well known activities. The $\gamma$ rays from the standard sources that were used, \isotope{Am}{241}, \isotope{Ba}{133}, \isotope{Cs}{137}, \isotope{Cd}{109}, and \isotope{Co}{57,60}, covered the energy range corresponding to the characteristic $\gamma$-ray lines of fluence monitor isotopes (411 to 1120 keV) and the \ta{180} decay (93.3 keV). Table~\ref{tbl:uncertainties} lists the characteristic $\gamma$-ray lines for every isotope that was irradiated in present work.
    
\begin{figure}[]
        \centering
        \includegraphics[width=1.0\linewidth]{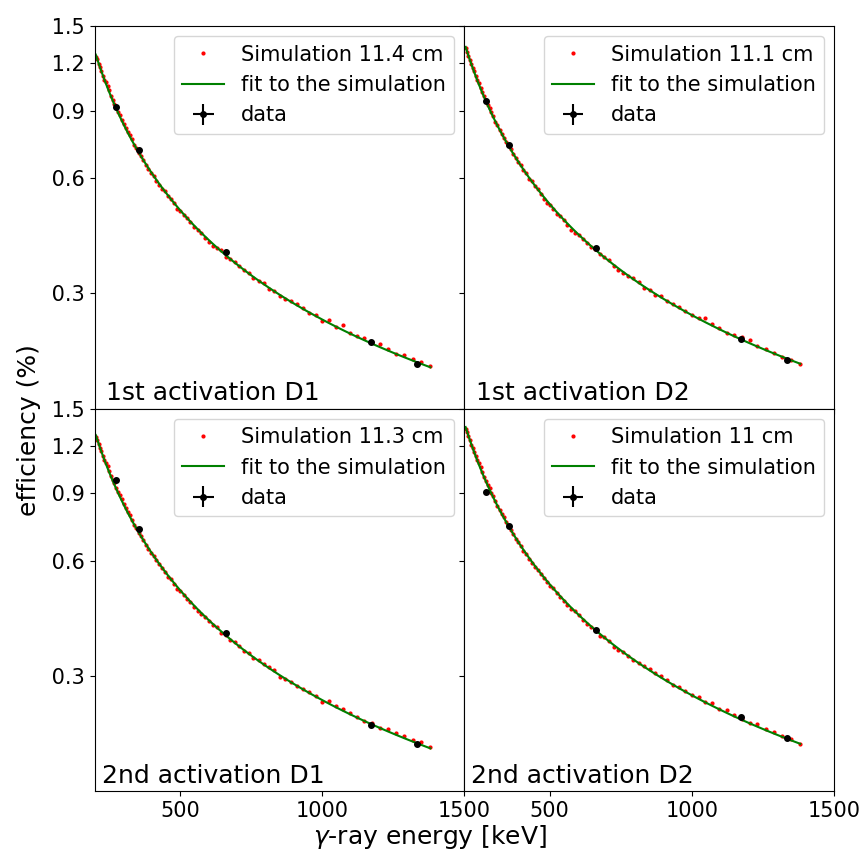}
        \caption{Detector efficiency as a function of energy for the two detectors D1 and D2 for the fluence monitors' spectroscopy setups after the 1st (no shield) and 2nd (Cd-shield) activations. Black circles show the measured values using radioactive sources. Error bars on the measurements are smaller than then marker size. Red dots are the simulated values obtained using GEANT3. The green line is the fit to the simulated values.}
        \label{fig:eff}
    \end{figure}

    For the higher energy range ($>$200 keV) corresponding to the $\gamma$-lines from the monitor isotopes, the detection efficiency can be described by the function,
    \begin{equation}
    \label{eqn:eff_fitfunc}
        \epsilon(E_{\gamma}) = A \exp[B + C\, ln(E_{\gamma}) + D (ln(E_{\gamma}))^2],
    \end{equation}
    where, A, B, C, and D are the free parameters. This function was used to fit the simulated $\epsilon(E_{\gamma})$ values for the monitors' spectroscopy setup, to obtain a continuous relationship between the energy and efficiency and evaluate the simulated efficiencies at the sources' $\gamma$ energies. Finally, the detector distances were determined by minimising the reduced-$\chi^2$(d), given by the following equation, 
    \begin{equation}
    \label{eqn:chisq}
        \chi^2(d) = \sum_{i} \frac{[\epsilon^{sim}(E_i,d) - \epsilon^{meas}(E_i)]^2}{[\Delta\epsilon^{meas}(E_i)]^2},
    \end{equation}
    where, $\epsilon^{sim}$ and $\epsilon^{meas}$ are simulated and measured efficiencies, respectively, $\Delta\epsilon^{meas}$ is the uncertainty on the measured values, d is the detector-to-source-holder distance in the simulation, and E$_i$ are the sources' $\gamma$ rays' energies. The simulated distance that resulted in the least $\chi^2$ was chosen as the best description of the real setup and was used to evaluate the detection efficiencies at the monitors' $\gamma$ energies. Figure \ref{fig:eff} shows the measured and best-match simulated efficiencies for the setup arrangements for fluence monitors' spectroscopy. The uncertainty in the detector distances was estimated to be 0.1~cm from the $\chi^2$ variation as a function of distance. This translated to a relative uncertainty of $<1$\% on the evaluated efficiencies.
    
    A similar procedure was applied to evaluate the detector distances and thus the efficiencies for the tantalum sample's spectroscopy setup. In this case, however, the fit function from equation~\ref{eqn:eff_fitfunc} could not be used as it is not valid in the low energy region. Instead, the detector distances were determined by direct comparison of the measured efficiencies with the simulation of the detector response at the energies corresponding the the radioactive sources used. The simulations of this setup also accounted for the attenuation from the absorber disc that was placed in front of the detectors to reduce the X-ray background (section~\ref{sec:setup}). At 93.3~keV, corresponding to the \ta{180} $\gamma$ ray energy, we obtained efficiency values for detector1/detector2 for the 1st and 2nd activations of 6.18$\times10^{-3}$/6.52$\times10^{-3}$ and 6.40$\times10^{-3}$/6.54$\times10^{-3}$, respectively. The relative uncertainly on these efficiency values were 2.3\%.

    More details of the effect of the absorber thickness and the distance between sample and detectors on $\gamma$-spectra can be found in Ref.~\cite{Dellmann21}.

    %=======================
    \subsection{Neutron fluence}
    %=======================
    \label{sec:fluence}
    
    For the first activation (without Cd shield), three monitor isotopes, i.e. \isotope{Zr}{94}, \isotope{Zr}{96}, and \isotope{Au}{197}, were sufficient to determine the fluence. The second activation (with Cd shield) required an additional monitor isotope with a higher sensitivity to the thermal neutrons due to the small thermal fluence. Scandium-45 has a high \(\sigma_{th}~to~I_{RI}\) ratio, and therefore, was used as a monitor in the second activation.
    
\begin{figure}[tp]
        \centering
        \includegraphics[width=1.0\linewidth]{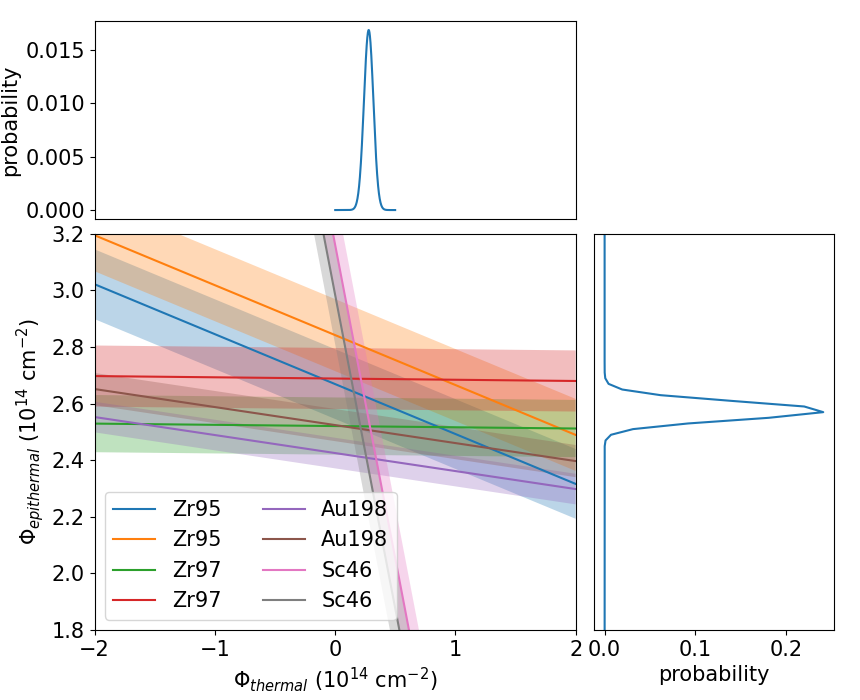}
        \caption{Epithermal vs thermal fluence plots as measured with different monitor isotopes (two samples of each) in the 2nd activation (with Cd shield). The top and the right plot give the probability distribution of the $\Phi_{th}$ and $\Phi_{epi}$ solutions, respectively.}
        \label{fig:fluence}
    \end{figure}

    For each activation, a set of linear equations for $\Phi_{th}$ and $\Phi_{epi}$ was obtained by inserting the activation ratios of the monitor isotopes (calculated using equation \ref{eqn:act_calc}) and the cross-section values (from table~\ref{tab:mon_iso}) in equation~\ref{eqn:act_gen}. Figure~\ref{fig:fluence} shows the eight linear equations for the four monitor isotopes (two samples each) that were used in the 2nd activation (without Cd shield). Since with several monitor reactions the equations are over-determined, the fluence values $\Phi_{th}$ and $\Phi_{epi}$ were obtained through combining the 2d probability distributions for each monitor and normalising the integral to 1. The details of the technique can be found in ref.~\cite{Dellmann22}.

    The fluence values for both activations from each of the detector measurements are listed in Table~\ref{tbl:fluxes_ta_act}. The excellent agreement between the values obtained from the two detectors is notable as each detector was treated independently in terms of dead time, efficiencies, and gamma-peak integration.
    
    {\renewcommand{\arraystretch}{1.1}
    \setlength{\extrarowheight}{1.0pt}
    \setlength{\tabcolsep}{3.5pt}
    \begin{table}[ht]
        \centering
        \begin{tabular}{|c|c|c|c|c|}
            \hline 
              & \multirow{2}{*}{\makecell{Activ-\\ation}} & $\Phi_{th}$ & $\Phi_{epi}$ & $N_{180}/N_{179}$\\
              & & ($10^{14}~cm^{-2}$) & ($10^{14}~cm^{-2}$) & ($10^{-7}$)\\
             \hline
             \hline
              \multirow{2}{*}{Det.1} & No Cd & 56.77 $\pm$ 1.39 & 2.62 $\pm$ 0.08 & 59.77 $\pm$ 3.06 \\ \cline{2-5}
               & Cd & 0.28 $\pm$ 0.04 & 2.60 $\pm$ 0.03 & 5.22 $\pm$ 0.39 \\
               \hline
              \multirow{2}{*}{Det.2} & No Cd & 56.25 $\pm$ 1.40 & 2.64 $\pm$ 0.08 & 58.52 $\pm$ 2.68 \\ \cline{2-5}
               & Cd & 0.28 $\pm$ 0.04 & 2.56 $\pm$ 0.03 & 5.70 $\pm$ 0.41 \\
               \hline
              \multirow{2}{*}{Avrg.} & No Cd & 56.49 $\pm$ 1.31 & 2.63 $\pm$ 0.08 & 59.15 $\pm$ 2.75 \\ \cline{2-5}
               & Cd & 0.28 $\pm$ 0.04 & 2.58 $\pm$ 0.03 & 5.46 $\pm$ 0.33 \\
               \hline
        \end{tabular}
        \caption{Fluence and tantalum activation values for the irradiations with and without Cd shielding as measured using the two detectors. The average values of the fluences were calculated using the averaged activation values of the monitor isotopes in the procedure described in section~\ref{sec:fluence}.}
        \label{tbl:fluxes_ta_act}
    \end{table}}
    
    %=======================
    \subsection{Cross-section calculations}
    %=======================
    
    The measured activation ratios ($N_{\ta{180}}/N_{\ta{179}}$) for the two activations of \ta{179} sample are given in last column of the table~\ref{tbl:fluxes_ta_act}. The activation values were calculated using equation~\ref{eqn:act_calc}.
    
    Using the fluence and activation ratio values from the two activations (table~\ref{tbl:fluxes_ta_act}) in equation~\ref{eqn:act_gen}, the following set of two linear equations was obtained, that was solved for thermal cross-section $\sigma_{th}$ and the resonance integral I$_{res}$ of \ta{179}.

    \begin{subequations}
    \begin{align}
         \sigma_{th}~\Phi_{th}~ + I_{res}~\Phi_{epi}~ &= \frac{N_{\ta{180}}}{N_{\ta{179}}}\label{eqn:act_ta1}\\
        \sigma_{th}~\Phi_{th}^{~Cd} + I_{res}~\Phi_{epi}^{~Cd} &= \left.\frac{N_{\ta{180}}}{N_{\ta{179}}}\right\vert_{Cd}\label{eqn:act_ta2}.
    \end{align}
    \end{subequations}
    
    The final cross-section results are shown in table~\ref{tbl:final_cs}.

    {
    \renewcommand{\arraystretch}{1.1}
    \setlength{\extrarowheight}{1.0pt}
    \setlength{\tabcolsep}{19pt}
    \begin{table}[h]
    \centering
    \begin{tabular}{|c|c|c|}
        \hline
        & $\sigma_{th}$ (b) & I$_{res}$ (b)\\
        \hline \hline
        Detector1 & 965 $\pm$ 63 & 1905 $\pm$ 167 \\
        \hline
        Detector2 & 939 $\pm$ 57 & 2125 $\pm$ 180\\
        \hline
        Average & 952 $\pm$ 57 & 2013 $\pm$ 148\\
        \hline
    \end{tabular}
    \caption{\ta{179}($n,\gamma$)\ta{180}$^{g.s.}$ reaction cross-section as measured in the present work. The average column values were calculated using the average fluence and activation values from tables~\ref{tbl:fluxes_ta_act}.}
    \label{tbl:final_cs}
    \end{table}
    }

    The uncertainties on the final cross-section values were calculated from the standard error propagation of the uncertainties on fluence and tantalum activation values that are listed in table~\ref{tbl:fluxes_ta_act}. The uncertainties on the fluence values (table~\ref{tbl:fluxes_ta_act}) were determined as the standard deviation in the probability distribution that were obtained in section~\ref{tbl:fluxes_ta_act} as shown in figure~\ref{fig:fluence}, which in turn were dependent on the uncertainties in the cross-sections of the monitors isotopes and their activation ratios. Given that the parameters in the activation ratio equation(\ref{eqn:act_calc}) were uncorrelated, their uncertainties were combined in the standard way to obtain the uncertainties on the activation ratios. Table~\ref{tbl:uncertainties} lists the relative uncertainties on the variables of equation~\ref{eqn:act_calc} for the tantalum sample and the fluence monitor samples numbered 4 (selected randomly).

    {
    \renewcommand{\arraystretch}{1.1}
    \setlength{\extrarowheight}{1.0pt}
    \setlength{\tabcolsep}{20.0pt}
    \begin{table*}[htb]
    \centering
    \begin{tabular}{|c||c|c|c|c|c|c|}
        \hline
        sample & Au-4 & \multicolumn{2}{c|}{Zr-4} & Sc-4 & \multicolumn{2}{c|}{Ta(No-Cd/Cd)} \\
        \hline
        \makecell{activated isotope} & \isotope{Au}{198} & \isotope{Zr}{95} & \isotope{Zr}{97} & \isotope{Sc}{46} & \multicolumn{2}{c|}{\isotope{Ta}{180}}\\
        \hline
        $E_{\gamma} (keV)$ & 411.8 & 724.2, 756.7 & 743.4 & 1120.5 & \multicolumn{2}{c|}{93.3} \\
        \hline
        $\Delta C_{\gamma}/C_{\gamma}$ & 0.68 & 3.22, 2.22 & 0.28 & 0.64 & 1.22 & 5.44 \\
        \hline
        $\Delta \epsilon_{\gamma}/\epsilon_{\gamma}$ & 1.00 & 1.00, 1.00 & 1.00 & 1.00 & 2.29 & 2.29 \\
        \hline
        $\Delta I_{\gamma}/I_{\gamma}$ & 0.10 & 0.50, 0.40 & 0.17 & 0.00 & \multicolumn{2}{c|}{3.55} \\
        \hline
        $\Delta f_{a}/f_{a}$ & 0.56 & 0.56 & 0.56 & 0.56 & 0.58 & 0.58 \\
        \hline
        $\Delta f_{w}/f_{w}$ & 0.00 & 0.00 & 0.19 & 0.00 & 0.30 & 0.30 \\
        \hline
        $\Delta f_{m}/f_{m}$ & 0.00 & 0.00 & 0.03 & 0.00 & 0.02 & 0.04 \\
        \hline
        $\Delta N_{tar}/N_{tar}$ & 0.17 & 0.14 & 0.14 & 0.65 & 2.55 & 2.58 \\
        \hline
        $\Delta R/R$ & 1.35 & 2.74 & 3.44 & 1.46 & 5.12 & 7.38 \\
        \hline
    \end{tabular}
    \caption{Relative uncertainties in \% on variables in the activation equation~\ref{eqn:act_calc}. The uncertainties on peak-counts, efficiencies, and activation ratio for both detectors were similar. The values shown here are for detector D1. The uncertainties for the fluence monitors listed here are for a typical sample per isotope (out of all that are listed in table~\ref{tab:mon_mass}).}
    \label{tbl:uncertainties}
    \end{table*}
    }

%======================= RESULTS and DISCUSSION =======================
\section{Summary and discussion}

In present work, we have measured the neutron-capture cross-section on \ta{179} at thermal and epithermal neutron energies via activation technique at TRIGA reactor, Mainz. The high neutron fluence at the reactor allowed for a cross-section measurement using only a 47~ng sample of \ta{179}. The target was produced via the \hf{180}(p,2n)\ta{179} reactions at MC40 cyclotron at the University of Birmingham, and the irradiated material was processed at PSI into a target.

We obtain a thermal cross section value of $952\pm57$ barns, and a resonance integral of I$_{res}$=2013 $\pm$ 148 barns (see table ~\ref{tbl:final_cs}). The only previous experimental data on the \ta{179}($n,\gamma$) reaction are by Schumann and K\"appeler \cite{Schumann99}, who obtained 932 $\pm$ 62 b and 1216$\pm$ 69 b for $\sigma_{th}$ and $I_{res}$, respectively. While the agreement for the thermal values is excellent, our resonance integral is 1.66 times higher. The reason for this large discrepancy is unknown. In both measurements, the same thickness (1~mm) of the Cd absorber was used. Any small inhomogeneities in the thickness would not lead to significantly different results in the two measurements, unless there is a strong resonance in the \ta{179}+n reaction at neutron energies around the cut off (0.1--1~eV)~\cite{Esch08}. However, at present, there is no experimental data on neutron resonances for this reaction. We therefore strongly encourage the investigation of the \ta{179}+n reaction in the eV-energy-regime at a time of flight facility.

The higher resonance integral may point to a higher neutron capture cross section in the stellar neutron energy (keV) range, which would lead to a higher production of $^{180}$Ta during the s-process. For any firm conclusions, however, a measurement of the cross section at stellar energies is required. In addition, cross section data on production of the isomeric state \ta{180m} are essential. This can be achieved by combining an activation measurement with a neutron time-of-flight study, the latter allowing determination of the total production of \ta{180}, i.e. \ta{180}$^{m}$+\ta{180}$^{g.s.}$.

\section*{Acknowledgements}
For the purpose of open access, the author has applied a Creative Commons Attribution (CC BY) licence to any Author Accepted Manuscript version arising from this submission. This work was supported by the UK Science and Facilities Council (ST/M006085/1, ST/V001043/1), European Research Council ERC-2015-StG Nr. 677497, and the HORIZON 2020 SANDA project Nr. 847552.

\end{document}